\documentclass[]{aastex63}
\usepackage{color}
\usepackage{bm}
\usepackage{cases}
\usepackage{multirow}
\usepackage{lineno}
\usepackage{textcomp}

\received{}
\revised{}
\accepted{}
\submitjournal{}
\shorttitle{}
\shortauthors{}
\begin{document}
	\title{Cosmological-model-independent determination  of Hubble constant from  fast radio bursts and Hubble parameter measurements}
	
	\author{Yang Liu}
	\affiliation{Department of Physics and Synergistic Innovation Center for Quantum Effects and Applications, Hunan Normal University, Changsha, Hunan 410081, China}
	
	\author{Hongwei  Yu}
	\affiliation{Department of Physics and Synergistic Innovation Center for Quantum Effects and Applications, Hunan Normal University, Changsha, Hunan 410081, China}
	\affiliation{
Institute of Interdisciplinary Studies, Hunan Normal University, Changsha, Hunan 410081, China	}

	\author{Puxun Wu}
	\affiliation{Department of Physics and Synergistic Innovation Center for Quantum Effects and Applications, Hunan Normal University, Changsha, Hunan 410081, China}
	\affiliation{
Institute of Interdisciplinary Studies, Hunan Normal University, Changsha, Hunan 410081, China	}

	\email{yangl@hunnu.edu.cn}
	%\correspondingauthor{Hongwei Yu}
	\email{hwyu@hunnu.edu.cn}
	%\correspondingauthor{Puxun Wu}
	\email{pxwu@hunnu.edu.cn}
	
	%\linenumbers

\begin{abstract}
We establish a cosmological-model-independent method to determine  the Hubble constant $H_0$ from the localized fast radio bursts (FRBs) and the Hubble parameter measurements from cosmic chronometers and obtain a first such determination $H_0=71\pm 3~\mathrm{km/s/Mpc}$,  with an uncertainty of 4\%,  from the eighteen  localized FRBs and nineteen  Hubble parameter measurements in the redshift range $0<z\leq0.66$.  
This value, which is independent of the cosmological model, is consistent with the results from the nearby type Ia supernovae (SN Ia) data calibrated by Cepheids  and the Planck cosmic microwave background radiation observations at the $1\sigma$  and 2$\sigma$ confidence level, respectively.  
Simulations show that the uncertainty of $H_0$ can be decreased to the level of that from the nearby SN Ia when mock data from 500 localized FRBs with 50 Hubble parameter measurements   in the redshift range of $0<z\leq1$ are used. Since  localized FRBs are expected to be detected in large quantities, our method will be able to give a reliable and more precise determination of $H_0$ in the very near future, which will help us to figure out  the possible origin of the Hubble constant disagreement. 

\end{abstract}

\section{introduction}
The cosmological constant $\Lambda$ plus cold dark matter ($\Lambda$CDM) model is the simplest cosmological model which fits the observational data very well. Based on the $\Lambda$CDM model, the Planck Cosmic Microwave Background (CMB) radiation observations give a tight constraint on the Hubble constant $H_0$:  $H_0=67.4\pm 0.5~\mathrm{km/s/Mpc}$ with an uncertainty about 0.7\%~\citep{Planck}. This result however has a more than $5\sigma$ deviation from $H_0= 73.04\pm1.04~\mathrm{km/s/Mpc}$ with the uncertainty being about 1.42\%,  which is determined by the nearby type Ia supernovae (SN Ia) calibrated by Cepheids~\citep{Riess2022}. Since these SN Ia, which are calibrated by using the distance ladder, are located in the very low-redshift region,  $H_0$ determined with them can be regarded as almost cosmological-model-independent.  The  disagreement of $H_0$ between two different observations  has became  the most serious crisis in modern cosmology~\citep{Riess2020,Valentino2021,Perivolaropoulos2022,Dainotti2021,Dainotti2022}, and it indicates  that the assumed $\Lambda$CDM model used to determine the Hubble constant may be inconsistent with our present Universe or there may be potentially unknown systematic errors in the observational data. It is worth noting however that several studies have not found any systematics that could explain the discrepancy~\citep{Feeney2018,Efstathiou2014,Riess2016,Cardona2017,Zhang2017,Follin2018,Riess2018a,Riess2018b}.  To precisely identify the possible origin of the $H_0$  disagreement, many other observational data are needed to constrain the Hubble constant. However, constraints from the vast majority of data usually depend on a pre-assumed cosmological model.

Undoubtedly,  a cosmological-model-independent determination  of the Hubble constant from observational data with a redshift region larger than that  of the nearby SN Ia may shed light on the possible origin of the $H_0$  disagreement. 
 In this letter,  we propose a cosmological-model-independent method to determine the value of $H_0$ from the fast radio bursts (FRBs) and the Hubble parameter $H(z)$ measurements from cosmic chronometers, 
and present a first determination of $H_0$ with  current such observational data that is independent of the cosmological model, i.e.,   $H_0$ ($H_0=71\pm 3~\mathrm{km/s/Mpc}$).  
Considering that a huge number of FRBs  will be detected in the near future as  more than one thousand  FRB events are expected very day~\citep{Xiao2021}, we expect to be able to achieve the precision of $H_0$ determination  at the level from the nearby SN Ia  with our method very soon. 

FRBs are a type of  frequently and mysteriously  transient signals  of millisecond duration with typical radiation frequency of $\sim$GHz~\citep{Lorimer2007,Petroff2019,Zhang2020,Zhang2022,CHIME/FRB Collaboration2021}. These signals are significantly dispersed by the ionized medium distributed along the path between the sources and the observer. The observed dispersion, quantified by the dispersion measure (DM), results mainly from the electromagnetic interaction between the signals and the free electrons in the intergalactic medium (IGM).  Since the effects of the free electrons on the signals are cumulative with the increase of traveling  distance of FRBs,
the DM-redshift relations of FRBs can be used for  cosmological purposes.  For example, they have been used  to determine the fraction of baryon mass in the IGM~\citep{Li2019,Lemos2022,Wei2019,Li2020}, to constrain the cosmological parameters \citep{Gao2014,Yang2016,Walters2018}, to explore the reionization history of our Universe \citep{Caleb2019,Linder2020,Beniamini2021,Bhattacharya2021,Hashimoto2021,Lau2021,Pagano2021,Heimersheim2022}, to measure the Hubble parameter~\citep{Wu2020}, to probe the interaction between dark energy and dark matter~\citep{Zhao2022a}, and so on.

FRBs have also been used to determine the  value of the Hubble constant~\citep{Li2018,Zhao2021,Hagstotz2022,James2022,Wu2022,Zhao2022b}.  
\cite{Hagstotz2022} and \cite{Wu2022} have obtained  constraints on $H_0$ of $62.3\pm9.1~\mathrm{km/s/Mpc}$ and $68.81^{+4.99}_{-4.33}~\mathrm{km/s/Mpc}$ by utilizing  nine and eighteen localized FRBs, respectively.  Using sixteen localized FRBs and sixty unlocalized FRBs, \cite{James2022} have achieved   $H_0=73^{-12}_{-8}~\mathrm{km/s/Mpc}$. These results  rely unavoidably on an assumed cosmological model, usually $\Lambda$CDM, since the theoretical value of $\mathrm{DM}$ used in these studies is model-dependent.  Interestingly,  the derivative of $\mathrm{DM}$ with respect to  the cosmic time $t$ is not dependent on any cosmological models  although $\mathrm{DM}$ is,  and moreover  it   is proportional to the Hubble constant squared. Therefore, 
the value of $H_0$ can be determined cosmological-model-independently if  the time variation of $\mathrm{DM}$   can be observed directly. However,   the time variation of $\mathrm{DM}$  due to the cosmic expansion is extremely weak, which is  about $-5.6\times 10^{-8} (1+z)^2~\mathrm{pc/cm^{3}/yr}$ with $z$ being redshift~\citep{Yang2017},  and thus it is very difficult to measure  directly. In this letter, we find a subtle way to avoid  this problem so as to obtain $H_0$ cosmological-model-independently with FRBs, i.e.,  we  propose  to derive  the time variation of $\mathrm{DM}$  through combining the redshift variation of $\mathrm{DM}$ of FRBs, which can be derived from the redshift distribution of FRBs,   and the Hubble parameter measurements.  Since the FRB and Hubble parameter data can be in the  higher redshift region,  our method contrasts with  those using  local measurements such as the Cepheid-calibrated SN Ia which can only be performed at very low redshifts.

 \section{Method}
As is well-known, the radio pulse will be dispersed when it travels through the ionized IGM, which will result in different arrival times for photons with different frequencies.
For two photons with the frequencies being $\nu_1$ and $\nu_2$ ($\nu_1<\nu_2$), respectively,  the delayed arrival time can be expressed as
\begin{eqnarray}
	\Delta t=\frac{e^{2}}{2 \pi m_{e} c}\left(\frac{1}{\nu_{1}^{2}}-\frac{1}{\nu_{2}^{2}}\right) \mathrm{DM}\,,
\end{eqnarray}
where $e$ and $m_e$ are the electron charge and mass, respectively, $c$  the speed of light, and  DM is defined as
\begin{eqnarray}
	\mathrm{DM}\equiv \int \frac{n_{e, z}}{1+z} d l \, .
\end{eqnarray}
Here $dl$ is an infinitesimal  proper length along the line of sight, and $n_{e,z}$  the number density of free electrons at redshift $z$.
Therefore,  DM carries the information of the distance and number density of free electrons.
The observed DM is a combination of  four  different components:
\begin{eqnarray}\label{eq:DMobs}
	\mathrm{DM_{obs}}=\mathrm{DM_{MW}^{ISM}}+\mathrm{DM_{MW}^{halo}}+\mathrm{DM_{IGM}}+\mathrm{DM_{host}}\ .
\end{eqnarray}
Here the subscript `$\mathrm{MW}$',  `$\mathrm{IGM}$'  and `$\mathrm{host}$' represent the contributions from the Milky Way, the intergalactic medium  and the host galaxy, respectively. 
The superscript `ISM' and `halo' denote the contributions from interstellar medium and halo of galaxy, respectively.
Among them, $\mathrm{DM_{IGM}}$ depends on the cosmological model.
However, since the IGM is inhomogeneous, we can only derive the average of $\mathrm{DM_{IGM}}$ theoretically \citep{Deng2014} 
\begin{eqnarray}\label{eq:mean_DM}
	\langle\mathrm{DM_{IGM}}\rangle&=&\frac{3cH_0\Omega_{b0}}{8\pi Gm_p}\int_{0}^{z} \frac{f_{\mathrm{IGM}}(z)f_e(z)(1+z)}{E(z)}dz\, ,
\end{eqnarray}
where $\Omega_{b0}$, $G$, $m_p$, and $f_\mathrm{IGM}(z)$ are the current baryon density parameter, the gravitational constant, the proton mass, and the fraction of baryon mass in the IGM, respectively, $E(z)=H(z)/H_0$ is the dimensionless Hubble parameter, which is given by the concrete cosmological model, and $f_e(z)=Y_\mathrm{H} f_{e,\mathrm{H}}(z)+\frac{1}{2}Y_\mathrm{He}f_{e,\mathrm{He}}(z)$ is the ratio of the number of free electrons to baryons in the IGM. Here $Y_\mathrm{H}\sim 3/4$ and $Y_\mathrm{He}\sim 1/4$ are the hydrogen (H) and helium (He) mass fractions, respectively, and $f_{e,\mathrm{H}}$ and $f_{e,\mathrm{He}}$ are the ionization fractions for H and He, respectively. 

It is worth noting that $\langle\mathrm{DM_{IGM}}\rangle$ is evolving with the cosmic expansion.
After applying  the relation $
	dz=-H_0 E(z) (1+z) dt$,
we obtain the derivative of $\langle\mathrm{DM_{IGM}}\rangle$ with respect to the cosmic time $t$
\begin{eqnarray}\label{eq:var_DM}
	\frac{d\langle\mathrm{DM_{IGM}}\rangle}{d t}
	&=&-\frac{3cH_0^2\Omega_{b0}}{8\pi Gm_p}f_\mathrm{IGM}(z)f_e(z)(1+z)^2\, ,
\end{eqnarray}
which indicates that the variation of $\langle\mathrm{DM_{IGM}}\rangle$ with time is  independent of the dimensionless Hubble parameter $E(z)$ and thus of any cosmological models. 
As a result, the value of the Hubble constant can be derived cosmological-model-independently  from
\begin{eqnarray}\label{eq:cal_H0}
	H_0&=&\left[-\frac{8 \pi G m_p }{3 c \Omega_{b0} f_\mathrm{IGM}(z) f_e(z)(1+z)^2}\frac{d\langle\mathrm{DM_{IGM}}\rangle}{d t}\right]^{\frac{1}{2}},
\end{eqnarray}
if one can obtain $d\langle\mathrm{DM_{IGM}}\rangle/d t$, and fix  $\Omega_{b0}$, $f_\mathrm{IGM}(z)$ and $f_e(z)$.
Since the variation of $\langle\mathrm{DM_{IGM}}\rangle$ from cosmic expansion is   about $-5.6\times 10^{-8} (1+z)^2~\mathrm{pc/cm^{3}/yr}$ \citep{Yang2017},  it is extremely weak and thus is very difficult to measure.
Fortunately,  $d\langle\mathrm{DM_{IGM}}\rangle/d t$ can be obtained as a product of $d\langle\mathrm{DM_{IGM}}\rangle/d z$ and $d z/dt$. The variation of $\langle\mathrm{DM_{IGM}}\rangle$ with time can be determined once both $d\langle\mathrm{DM_{IGM}}\rangle/d z$ and $d z/dt$  are known. The $dz/dt$  factor can be found directly from the Hubble parameter measurements from cosmic chronometers. If we work out the relation of $\langle\mathrm{DM_{IGM}}\rangle$ with $z$ from the observational data of FRBs,   $d\langle\mathrm{DM_{IGM}}\rangle/d z$ at the redshifts of the Hubble parameter data points can be derived and then $d\langle\mathrm{DM_{IGM}}\rangle/d t$ at the same redshifts can be obtained. 
For example,   employing  a continuous piecewise linear function to approximate the $\langle\mathrm{DM_{IGM}}\rangle-z$ relation, we have
\begin{eqnarray}\label{eq:linear}
	\langle\mathrm{DM_{IGM}}\rangle(z)=\frac{\langle\mathrm{DM_{IGM}}\rangle_{i+1}-\langle\mathrm{DM_{IGM}}\rangle_i}{z_{i+1}-z_i} \left(z-z_i\right)+\langle\mathrm{DM_{IGM}}\rangle_i \, ,
\end{eqnarray}
after dividing uniformly the redshift range of FRB samples into $n$ bins with $n+1$ control points $z_i$. Here $\langle\mathrm{DM_{IGM}}\rangle_i$ is the undetermined dispersion measure at $z_i$, and $z_1=0$ is fixed. 
For the $n=1$ case, this function reduces to a linear function, and $z_2$ is the maximum  redshift  of  FRB samples. 
Thus, we can fix  $\langle\mathrm{DM_{IGM}}\rangle_1=0$, and obtain $\langle\mathrm{DM_{IGM}}\rangle_2$ from  the observed $\langle\mathrm{DM_{IGM}}\rangle$ by using the  Markov Chain Monte Carlo (MCMC) method. 
Taking the derivative of Eq.~(\ref{eq:linear}) with respect to $z$ yields  the $d\langle\mathrm{DM_{IGM}}\rangle/d z-z$ relation, from which the values of $d\langle\mathrm{DM_{IGM}}\rangle/d t$ at the redshifts of the Hubble parameter measurements can be obtained after using the $H(z)$ data.  Therefore, using Eq.~(\ref{eq:cal_H0}), we can constrain the Hubble constant.

\section{Data and results}
We will use the latest localized FRBs  data and the $H(z)$ data to determine  $d\langle\mathrm{DM_{IGM}}\rangle/d t$.
Our FRB samples are compiled in~\citep{Wu2022}, which contain eighteen FRBs within the redshift range of $z\in (0,0.66]$~\citep{Chatterjee2017,Bannister2019,Prochaska2019,Ravi2019,Bhandari2020,Heintz2020,Law2020,Marcote2020,Bhardwaj2021,Chittidi2021,Bhandari2022}.
The number of the latest $H(z)$ data is 32, spanning redshifts from 0.07 to 1.965~\citep{Simon2005,Stern2010,Moresco2012,Cong2014,Moresco2015,Moresco2016,Ratsimbazafy2017,Borghi2022}, which are measured by using the cosmic chronometric technique~\citep{Jimenez2002}. Here we select only nineteen $H(z)$ data that fall in the  redshift range of the FRB samples.

Since DM$_{\mathrm{obs}}$ is released for the FRB data,  we use Eq.~(\ref{eq:DMobs}) to extract the extragalactic DM by deducting the contribution from the Milky Way in $\mathrm{DM_{obs}}$:
\begin{eqnarray}
	\mathrm{DM_{ext}^{obs}}=\mathrm{DM_{obs}}-\mathrm{DM_{MW}^{ISM}}-\mathrm{DM_{MW}^{halo}}-\Delta\mathrm{DM_{IGM}} ,
\end{eqnarray}
where $\Delta\mathrm{DM_{IGM}}$ represents the contribution of fluctuations of the electron density in the IGM, which is assumed to obey a normal distribution $\mathcal{N}(0,\sigma_\mathrm{\Delta DM_{IGM}})$  since the fluctuations in the electron density along the line of sight can be approximated by a Gaussian distribution~\citep{Jaroszynski2019,Macquart2020,Zhang2021}. The $\mathrm{DM_{MW}^{ISM}}$ can be obtained from the current electron-density model of the Milky Way~\citep{Yao2017}, and $\mathrm{DM_{MW}^{halo}}$ is assumed as $65~\mathrm{pc/cm^{3}}$~\citep{Prochaska2019b}. 
Then the uncertainty $\sigma_\mathrm{DM_{ext}}$ of $\mathrm{DM_{ext}^{obs}}$ has the form
\begin{eqnarray}\label{eq:err_DM}	\sigma_\mathrm{DM_{ext}}^2=\sigma_\mathrm{DM_{obs}}^{2}+\sigma_\mathrm{DM_{MW}}^{2}+\sigma_\mathrm{\Delta DM_{IGM}}^{2}.
\end{eqnarray}
Here $\sigma_\mathrm{DM_{obs}}$ is given by the observation, $\sigma_\mathrm{\Delta DM_{IGM}} $ is the uncertainty of $\Delta\mathrm{DM_{IGM}}$, which is estimated through the approximation $\sigma_\mathrm{\Delta DM_{IGM}}/\langle\mathrm{DM_{IGM}}\rangle=20 \%/\sqrt{z}$ given in~\citep{Kumar2019}, and the uncertainty of $\mathrm{DM_{MW}}$ (containing the uncertainties of ISM and halo) is taken to be $54~\mathrm{pc/cm^{3}}$~\citep{Heimersheim2022}.
Apparently,  the theoretical value of $\mathrm{DM_{ext}}$ can be expressed as 
\begin{eqnarray}\label{eq:chi}
\mathrm{DM_{ext}^{th}}=	\langle \mathrm{DM_{IGM}}\rangle+\mathrm{DM_{host}}\ .\end{eqnarray}
However, the contribution of the host galaxy ($\mathrm{DM_{host}}$) in Eq.~(\ref{eq:chi}) is not easy to determine since we do not know it very well. Here, 
we follow \citep{Macquart2020,Zhang2020b} to consider a prior log-normal distribution of $\mathrm{DM_{host}}$ \begin{eqnarray}\label{eq:phost}
	P_\mathrm{host}(\mathrm{DM_{host}})=\frac{1}{\sqrt{2\pi\sigma^2}\mathrm{DM_{host}}}\exp\left[ -\frac{(\ln \mathrm{DM_{host}}-\mu)}{2\sigma^2} \right].
\end{eqnarray}
For this log-normal distribution,  the median and variance of $\mathrm{DM_{host}}$ are $e^\mu$ and $e^{2\mu+\sigma^2}(e^{\sigma^2}-1)$, respectively.
In~\citep{Zhang2020b}, the value of the median $e^\mu$ is assumed to be redshift-evolutionary:  $e^\mu \equiv A(1+z)^{\alpha}$ with $A$ and $\alpha$ being two constants. 
Once the allowed regions of $A$ and $\alpha$ are    determined from $\mathrm{DM_{ext}^{obs}}$, their best fitting values and uncertainties give the median and variation of $\mathrm{DM_{host}}$, respectively. Before  running the MCMC to constrain all free parameters,   we need to set the prior regions of  $A$ and $\alpha$, which are obtained by using the IllustrisTNG simulation~\citep{Zhang2020b}.
Moreover, as what was done in~\citep{Zhang2020b}, we place the host galaxies of FRBs into three types: (I) The repeating FRBs in a dwarf galaxy like the FRB 121102, (II) the repeating FRBs in a spiral galaxy like the FRB 180916, and (III) the non-repeating FRBs. 
Thus, we  set $\left\{ A_1,\;\alpha_1,\;A_2,\;\alpha_2,\;A_3,\;\alpha_3 \right\}$ as free parameters to describe the values of FRB' $\mathrm{DM_{host}}$ in Eq. (\ref{eq:chi}). These six parameters will be fitted simultaneously with the coefficients in Eq.~(\ref{eq:linear}), and  are marginalized in the subsequent analysis.

\begin{figure}
	\centering
	\includegraphics[width=.7\textwidth]{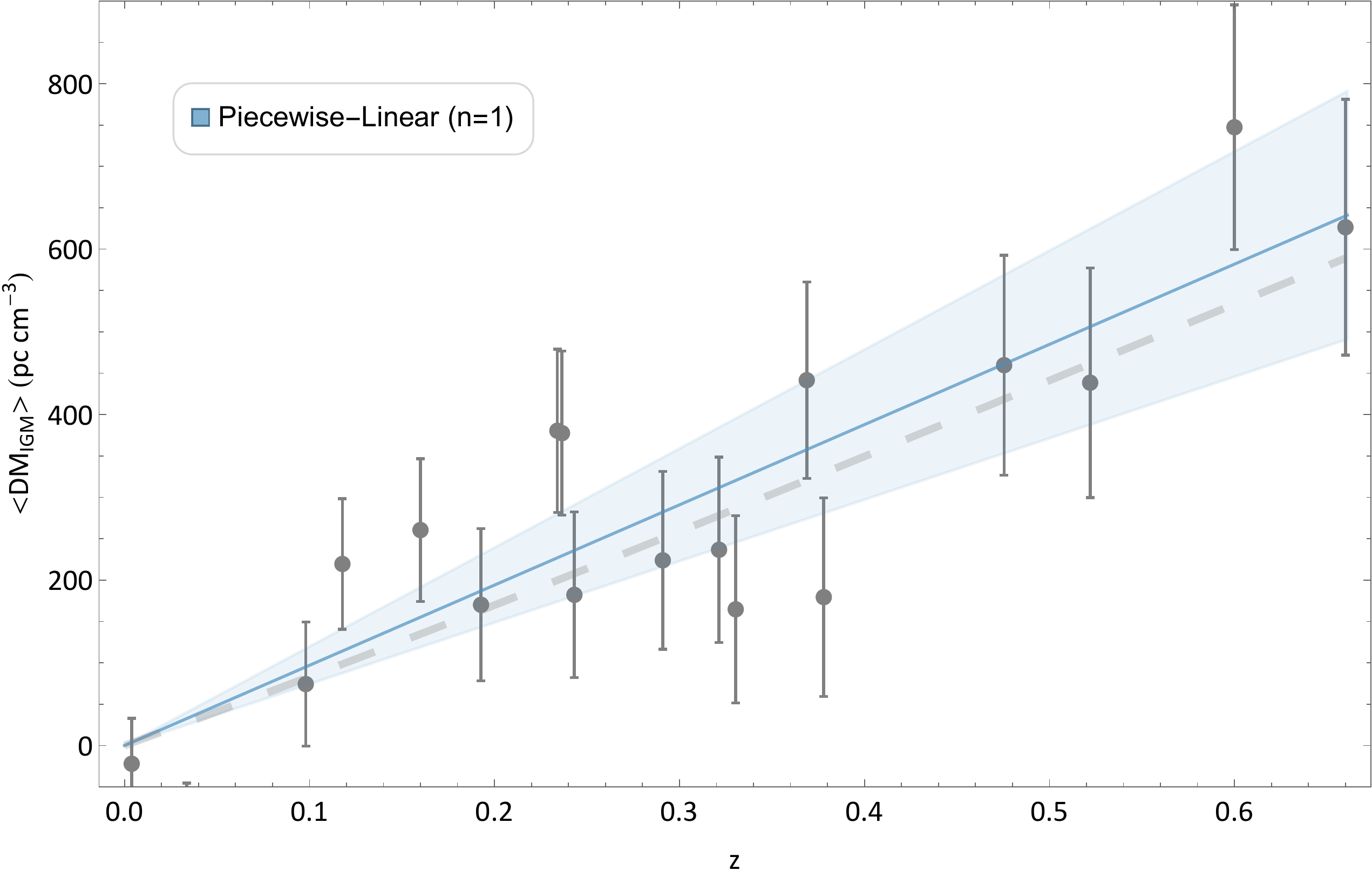}
	\caption{ 
		The $\langle\mathrm{DM_{IGM}}\rangle-z$ relation for the $n=1$ piecewise linear  function (blue line). The shadow region denotes the 1$\sigma$ uncertainty. 
		The gray points are  eighteen   $\langle\mathrm{DM_{IGM}} \rangle$ data samples.
		The dashed line is the theoretical value of $\langle\mathrm{DM_{IGM}} \rangle$ based on the $\Lambda$CDM model.
		\label{fig1}
	}
\end{figure}

\begin{figure}
	\centering
	\includegraphics[width=.6\textwidth]{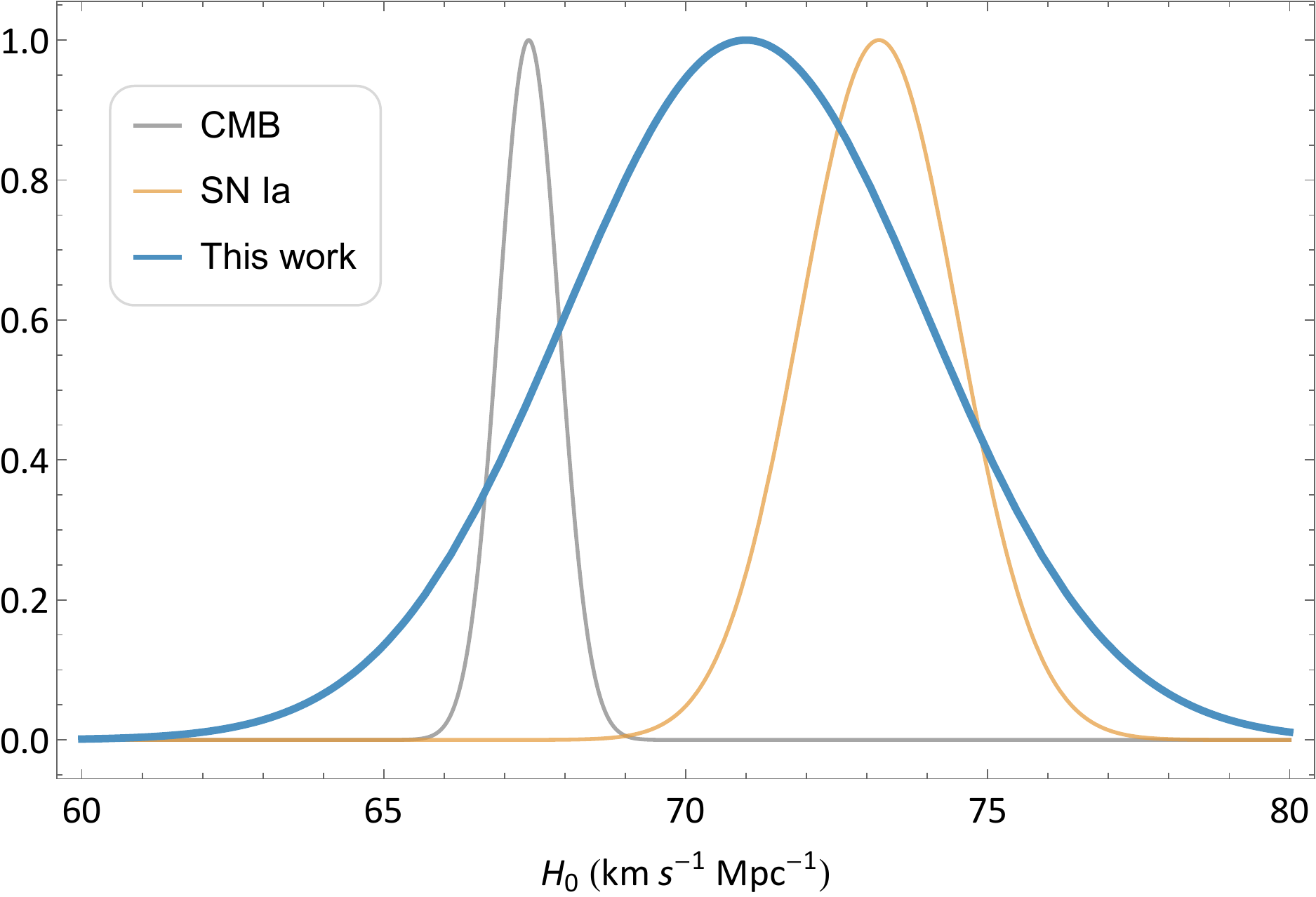}
	\caption{ 
		The constraints on $H_0$.
		The gray and orange lines represent the results from the Planck CMB observations and the nearby SN Ia data, respectively. The blue line shows our result.
		\label{fig2}
	}
\end{figure}

For the piecewise linear function  with $n=1$, we obtain $\langle\mathrm{DM_{IGM}}\rangle_2=641^{+144}_{-154}~\mathrm{pc/cm^{3}}$ from the eighteen FRB data points.  Figure \ref{fig1} shows the approximate $\langle \mathrm{DM_{IGM}}\rangle-z$ relation.
Combining nineteen Hubble parameter data with the values of $d\langle\mathrm{DM_{IGM}}\rangle/d z$  at the redshifts of the $H(z)$ data, we obtain nineteen $d\langle\mathrm{DM_{IGM}}\rangle/d t$ data points. 
To further calculate the value of $H_0$,  we fix  $\Omega_{b0}=0.0487\pm0.0005$~\citep{DES Collaboration2022}.
Due to the lack of evidence for the evolution of $f_\mathrm{IGM}$ over the redshift range covered by the FRB sample,  we adopt, in our analysis, $f_\mathrm{IGM}(z)=0.84^{+0.16}_{-0.22}$, which is determined by a cosmological-insensitive method~\citep{Li2020}. 
Since the H and He are fully ionized at $z<3$ \citep{Meiksin2009,Becker2011}, \textit{i.e.}, $f_{e,\mathrm{H}}=f_{e,\mathrm{He}}=1$, we set $f_e(z)=7/8$ in our analysis.
Finally, nineteen $H_0$ are derived from Eq.~(\ref{eq:cal_H0}).
Using the minimum $\chi^2$ method, we arrive at  a constraint on $H_0$:  $H_0=71\pm 3~ \mathrm{km/s/Mpc}$ with an uncertainty of 4\%. Figure \ref{fig2} shows a comparison between our result with those obtained by the Planck CMB observations~\citep{Planck} and the nearby SN Ia data~\citep{Riess2022}.  
Our result is consistent with that from nearby SN Ia at the 1$\sigma$ confidence level (CL), but that from  the CMB observations only at the $2\sigma$ CL.

 Since  $\Omega_{b0}=0.0487\pm0.0005$~\citep{DES Collaboration2022}, which is used in above  analysis,  depends on the $\Lambda$CDM model, to study the effect of this model-dependent value on our result, 
we also consider  $\Omega_{b0}=0.048\pm0.001$ from the $w$CDM model~\citep{DES Collaboration2022}, and obtain $H_0=71\pm 3~\mathrm{km/s/Mpc}$.
This result is consistent very well with that in the case of $\Omega_{b0}=0.0487\pm0.0005$.
To further investigate the effect of the uncertainty of  $\Omega_{b0}$ on our result, we choose a value of $\Omega_{b0}$ with a larger uncertainty:   $\Omega_{b0}=0.0487\pm 0.02$, and achieve  $H_0= 71\pm 5~\mathrm{km/s/Mpc}$ with $7$\% uncertainty. Apparently, when the uncertainty of $\Omega_{b0}$ increases 40 times (from $0.0005$ to $0.02$), the uncertainty of $H_0$ only increases $3\%$, which indicates that the precision of $H_0$ does not depend  sensitively on the  uncertainty of $\Omega_{b0}$. Thus, we conclude that the value of $H_0$ from our method is insensitive to  $\Omega_{b0}$. 

Let us now examine whether the adoption of the $n=1$ piecewise linear function leads to some bias in our results.  For this purpose, we perform a further analyis by using the $n=2$ piecewise linear function   and the quadratic polynomial function ($\langle\mathrm{DM_{IGM}}\rangle(z)=A z+B z^2$) to approximate the $\langle\mathrm{DM_{IGM}}\rangle-z$ relation,  where A and B are two constants. The constraints on the Hubble constant are $H_0=71\pm 5~\mathrm{km/s/Mpc}$ and  $H_0=71\pm 4~\mathrm{km/s/Mpc}$  for  the $n=2$ piecewise linear function   and the quadratic polynomial, respectively.    Apparently,  the constraints on $H_0$ are very well  consistent with each other  for three different approximations. Thus, we can conclude that the $H_0$ results are almost independent of the functions chosen to approximate the redshift evolution of $\langle\mathrm{DM_{IGM}}\rangle(z)$.

Let us note that  our results are slightly tighter than $H_0=75.7^{+4.5}_{-4.4}~\mathrm{km/s/Mpc}$  obtained model-independently from four strong gravitational lensing systems  and  SN Ia~\citep{Collett2019}, which is later improved to   $H_0=72.8^{+1.6}_{-1.7}~\mathrm{km/s/Mpc}$ when six strong gravitational lensing systems are used~\citep{Liao2020}. These cosmological-model-independent results from strong lensing systems and SN Ia are consistent with $72.5^{+2.1}_{-2.3}~\mathrm{km/s/Mpc}$~\citep{Birrer2019} and  $73.3^{+1.7}_{-1.8}~\mathrm{km/s/Mpc}$~\citep{Wong2020} obtained before from the four and six lensing systems respectively with an assumed spatially flat $\Lambda$CDM model.  Furthermore, 
simulations show that 400 lensing systems  can constrain $H_0$ model-independently with an uncertainty at the level of the nearby SN Ia~\citep{Collett2019}.

To see how many FRBs are needed for a  determination  $H_0$ as precise as of the level of the nearby SN Ia, we need to use the Monte Carlo simulation.
During simulation, we choose the spatially flat $\Lambda$CDM model with $H_0=73.04~\mathrm{ km/s/Mpc}$ and $\Omega_\mathrm{m0}=0.3$ to be the fiducial model and  set $\Omega_{b0}=0.0487$, $f_\mathrm{IGM}=0.84$, and $f_e(z)=7/8$. 
Here we randomly sample $\mathrm{DM_{ext}^{sim}}$ as the  observed quantity, which is obtained from $\mathrm{DM_{IGM}^{sim}}+\mathrm{DM_{host}^{sim}}$. 
The redshift distribution of FRBs is assumed to be $P(z)\propto z^2 \exp (-7z)$ in the redshift range $0<z\leq1$ \citep{Hagstotz2022}. 
At the mock redshift $z$, the fiducial value of $\langle \mathrm{DM_{IGM}^{fid}} \rangle$ can be calculated from Eq.~(\ref{eq:mean_DM}).  
The $\mathrm{DM_{IGM}^{sim}}$ is sampled from $\mathcal{N}\left(\langle \mathrm{DM_{IGM}^{fid}} \rangle, \sigma_\mathrm{\Delta DM_{IGM}}  \right)$ with $\sigma_\mathrm{\Delta DM_{IGM}}=20\% \langle \mathrm{DM_{IGM}^{fid}} \rangle/\sqrt{z}$~\citep{Kumar2019}. 
 $\mathrm{DM_{host}^{sim}}$ is simulated by using  the distribution $P_\mathrm{host}(\mathrm{DM_{host}})$ given in Eq. (\ref{eq:phost}). The type of the host galaxy is chosen randomly as  one of the three different types, and  the prior values of the parameters $\mu$ and $\sigma$ are set from the IllustrisTNG simulation~\citep{Zhang2020b}.
Meanwhile, we also sample the $H(z)$ data with a uniform distribution at $0<z\leq1$ following \citep{Ma2011}.  We simulate 500 FRBs and 50 $H(z)$ data. 
The larger number of data set prompts us to use the piecewise linear function with $n=2$ to approximate $\langle\mathrm{DM_{IGM}}\rangle-z$ relation.
To minimize randomness and to ensure that the final constraint result is unbiased, we repeat the above steps 100 times and finally obtain $H_0=73.7\pm1.4~\mathrm{km/s/Mpc}$. This indicates that the uncertainty of $H_0$ can be decreased to 1.9\% from mock data, which is almost of the same precision as that  from the nearby SN Ia (1.42\%).

A high  precision determination of $H_0$ from FRBs is expected soon since a large number of localized FRBs will be detected in the next few years.  
This is because  there are  huge number of FRB events every day and the detection ability is being  improved rapidly. 
The running  and upcoming radio  telescopes and surveys include the Swinburne University of Technology's digital backend for the Molonglo Observatory Synthesis Telescope array (UTMOST)~\citep{Caleb2016}, the Canadian Hydrogen Intensity Mapping Experiment (CHIME)~\citep{CHIME/FRB Collaboration2021,Bandura2014}, the Hydrogen Intensity and Real-time Analysis eXperiment (HIRAX)~\citep{Newburgh2016}, the Five-hundred-meter Aperture Spherical Telescope (FAST)~\citep{LiDi2018}, and  the Square Kilometre Array (SKA) project~\citep{Macquart2015,Fialkov2017}.

%{\it Conclusion.} 
\section{Conclusion}
The disagreement between the measurements of  the Hubble constant from the CMB observations and the nearby SN Ia data has became one of the pressing challenges in modern cosmology.
A cosmological-model-independent method to determine the value of $H_0$ from the data in the redshift region larger than that of the nearby SN Ia  may serve as a probe to the possible origin of $H_0$ disagreement.  In this letter,  we establish  a feasible way to cosmological-model-independently constrain  $H_0$ by combining the variation of $\langle\mathrm{DM_{IGM}}\rangle$ with the redshift of FRBs and the Hubble parameter measurements, and 
obtain a first such determination $H_0=71\pm 3~\mathrm{km/s/Mpc}$  with data  from the eighteen  localized FRBs and nineteen  Hubble parameter measurements in the redshift range $0<z\leq0.66$. 
Remarkably, this  value, which is independent of the cosmological model, lies in the middle of the results from CMB observations and the nearby SN Ia data,  and it  is consistent with those from the nearby SN Ia data   and  the CMB observations at the $1\sigma$ and $2\sigma$ CL, respectively.  The  uncertainty of our result is much less than what were obtained from FRBs in the framework of $\Lambda$CDM model~\citep{Hagstotz2022,Wu2022,James2022}.

However, as our result has large uncertainty, it does not show  significant statistic evidence for preferring the result from the nearby SN Ia data. 
Through the Monte Carlo simulation, we further investigate how many FRBs and $H(z)$ measurements are needed to more precisely determine the value of $H_0$.
We find that the uncertainty of $H_0$ from mock 500 localized FRBs and 50 $H(z)$ data at $0<z\leq1$ can be decreased to $1.9\%$, which is of the same level as that  from the nearby SN Ia data.
Since  localized FRBs  are expected to be detected in large quantities, the method established in this paper will be able to  give a reliable and more precise determination of $H_0$ in the very near future, which will help us to figure out the possible origin of the Hubble constant disagreement.

\acknowledgments
%{\it Acknowledgments.}
%We appreciate very much the insightful comments and helpful suggestions by anonymous referees.
This work was supported in part by the NSFC under Grant Nos. 12275080, 12075084,   11805063, and 12073069.   %   by the National Key Research and Development Program of China Grant No.2020YFC2201502, and by the Science and Technology Innovation Plan of Hunan province under Grant No. 2017XK2019.
%\end{acknowledgments}

\end{document}